# A Search for the Quark-Hadron Phase Transition in pp collisions at $\sqrt{s}$ = 13 TeV using UrQMD model

Swarnapratim Bhattacharyya

**Department of Physics, New Alipore College, L Block, New Alipore, Kolkata 700053, India**

Email: swarna_pratim@yahoo.com

## Abstract

Charged particle multiplicity fluctuations in pp collisions at $\sqrt{s}$ = 13 TeV have been studied by the method of scaled factorial moment for the minimum bias events generated by Ultra Relativistic Quantum Molecular Dynamics (UrQMD) model in the pseudo-rapidity (η), azimuthal angle (ϕ) and two dimensional anisotropic (η − ϕ) phase space. Strong intermittent type of fluctuations have been observed for the UrQMD simulated data. Systematic studies of intermittent fluctuations in terms of scaled factorial moments was utilized to extract the anomalous fractal dimension in the pseudo-rapidity (η), azimuthal angle (ϕ) and two dimensional anisotropic (η − ϕ) phase space. Search for the quark-hadron phase transition in the framework of Ginzburg-Landau theory of second-order phase transition by the analysis with the factorial moment method has also been performed.

# Introduction

Heavy ion physics aims to understand the physics of extended hadronic matter under extremes of energy density and temperature only reachable in the hot early universe, by colliding heavy nuclei at relativistic and ultra-relativistic energies. Such experiments yield a rich variety of interesting data which can provide valuable information on the spatiotemporal development of multiparticle production process. It was noted even before the advent of Quantum Chromodynamics (QCD) as the underlying theory of strongly interacting elementary particles that nuclear matter cannot exist as hadrons at an arbitrarily high temperature or density. It has been found from the theory of lattice quantum chromodynamics (QCD) that at zero net baryon density at a critical temperature of $T_c \sim$ 170 MeV and energy density $\varepsilon_c \sim$ 1 GeV fm$^{-3}$ a color deconfined and a chirally restored quark-gluon-plasma (QGP) phase is formed. These energy densities may be attained in relativistic and ultra relativistic heavy-ion collisions where it is believed that a dense system of quarks and gluons is created. The system undergoes rapid collective expansion before the partons hadronize and eventually decouple. The primary goal of relativistic heavy-ion collisions is to create and study this deconfined state in the laboratory. At very high energies at the LHC and the RHIC, the deconfined phase exist for a brief time before it cools down and rehadronizes and so we cannot study it directly. Among the various signals to understand the presence of this QGP state, an important one is the fluctuations of the produced particles. Lattice QCD predicts large fluctuations being associated with the system undergoing phase transition. Such fluctuation is much larger than the statistical fluctuations arising due to the finiteness of the particles produced in a collision. Studies of particle density fluctuations have prompted considerable advances in this area of research. Among the various method of studying particle density fluctuations, the method of moment was found to be most effective. But the presence of statistical fluctuations creates a lot of problems and the problems become more intense when the scale dependence of fluctuations is investigated. To gather any meaningful information about the particle production mechanism, it is therefore important to disentangle the statistical noise arising due to the finite number of produced particles in the final state. In order to overcome this difficulty the method of factorial moments was proposed by Bialas and Peschanski [1-2]. This method is capable of filtering out the statistical noise. The scaled factorial moments of integer orders capable of detecting and characterizing non-statistical density fluctuations, are found to depend on the phase-space resolution ($\delta x$) obeying a power law of the type $\langle F_q \rangle \propto (\delta x)^{-\alpha_q}$. $\alpha_q$ is known as intermittency index. This power-law scaling behavior of scaled factorial moments ($F_q$) is known as "intermittency" [1-2]. Application of factorial moment method to JACEE events [3] not only confirmed the extrication of statistical fluctuations but also supplied evidence for a power law dependence of fluctuations on the bin size. The observation of the power law behavior of the scaled factorial moments with decreasing size of the phase space interval indicates the presence of self similar fluctuations in multiparticle production in high energy interactions [4]. A detail review on intermittent fluctuations can be found from [4]. The proposal to look for intermittency also prompts a thorough study of fractality. The fractal structure is a natural consequence of cascading mechanism prevailing in the multiparticle production. The self-similarity observed in the power law dependence of scaled factorial moments reveals a connection between intermittency and fractality. The scaled factorial moment technique empowers us to correlate the intermittency exponent $\alpha_q$ to the anomalous fractal dimension $d_q$ by the relation as given in [5] $d_q = \frac{\alpha_q}{q-1}$. Significance of anomalous fractal dimension lies in the fact that it reflects the nature of fractal structure of multiparticle production. From the dependence of anomalous fractal dimension on the order q a distinguishable characterization of fractality is possible [5]. Increase of anomalous fractal dimension $d_q$ with the order of moment q signals the presence of multifractality. On the other hand if $d_q$ remains constant with the increase of order of moment q monofractality occurs.

Fractal nature of multiparticle production reflects the possibility of quark-hadron phase transition. R.C. Hwa [6] pointed out that the study of multiplicity fluctuations of hadrons produced in high energy heavy ion interactions can be used as a means to detect evidence of a quark-hadron phase transition. In analogy with the photo count problem at the onset of lasing in non linear optics, the coherent state description can be used in the frame work of Ginzburg-Landau theory. A quantity $\beta_q$ in terms of the ratio of higher order anomalous fractal dimension ($d_q$) to the second order anomalous fractal dimension ($d_2$) can be defined by the following relation

$$\beta_q = \frac{d_q}{d_2}(q-1)\ldots\ldots\ldots\ldots\ldots\ldots(1)$$

According to Ginzburg-Landau model [7] $\beta_q$ is related with (q-1) by the relation

$$\beta_q = (q-1)^\nu \ldots\ldots\ldots\ldots\ldots\ldots (2).$$

The quantity $\nu$ is the indicator of phase transition study.

The relation (1) and relation (2) are found to be valid for all systems which can be described by the Ginzburg-Landau (GL) theory and also is independent of the underlying dimension of the parameters of the model [7]. The quantity $\nu$ is a universal quantity. If the value of the scaling exponent $\nu$ is equal to or close enough (within the experimental error) to 1.304 for the experimental data then a quark-hadron phase transition is expected [7]. If the measured value of $\nu$ is different from the critical value 1.304 considering the experimental errors then the possibility of the quark-hadron phase transition has been faded. The study of quark-hadron phase transition in terms of the critical exponent $\nu$ is of particular importance in high energy collisions where neither the transition temperature nor the other important parameters related to phase transitions are known. If a signature of quark-hadron phase transition depends on the details of the heavy-ion collisions e.g. nuclear sizes, collision energy, transverse energy, etc., then even after the system has passed the thresholds for the creation of QGP, such a signature is likely to be sensitive to this theory. It may be mentioned here that the investigation of phase transition is quite common in the field of statistical physics. This analogy has been extended in the field of high energy nucleus-nucleus collisions. In the original formulation of Ginzburg-Landau theory of phase transition, systematic studies of intermittent fluctuations in terms of scaled factorial moments was utilized to extract the anomalous fractal dimension and hence the critical coefficient was calculated [7].

**Goal of the Present Study**

Enormous experimental efforts at the Super Proton Synchrotron (SPS) at CERN [8], the Relativistic Heavy Ion Collider (RHIC) [9-15] at BNL, and the Large Hadron Collider (LHC) [16-19] at CERN proved the experimental evidence for the existence of the QGP state in nucleus-nucleus collisions at relativistic and ultra-relativistic energies over a wide range of energy and centrality. Various probes were used to determine the features of the phase diagram of the nuclear matter. However in the proton-proton (pp) collisions which are considered as the reference system for heavy ion collisions, such existence of QGP state may not be expected.

Along with the study of heavy ion collisions, a thorough understanding of proton-proton (pp) collisions at ultra relativistic energies is also necessary both as input to detailed theoretical models of strong interactions, and as a baseline for understanding the nucleus–nucleus collisions at the RHIC and at the LHC energies. Soft particle production from ultra relativistic pp collisions is sensitive to the flavor distribution within the proton, quark hadronization and baryon number transport. The measurement of charged particle transverse momentum spectra in pp collisions serves as a crucial reference for particle spectra in nucleus-nucleus collisions. A proton-proton reference spectrum is needed for nucleus-nucleus collisions to investigate possible initial-state effects in the collision. It is therefore important to investigate the pp collisions so that the properties of this system can be measured and understood over a wide range of energies. Due to early thermalization, it also remains doubtful if the matter formed in such collisions exhibits collective like behavior as observed in heavy-ion collisions [20]. With the advent of modern particle accelerator at the LHC for the proton-proton collision system, the multiplicity of pp collision events becomes comparable to that of high energy nucleus-nucleus collisions [21]. This encourages the scientists to study the pp collisions in some more details and systematic studies on various observable of high-multiplicity pp events have become essential for a better understanding of the dynamics of such collisions.

In this work, an attempt has been made with Ultra Relativistic Quantum Molecular Dynamics (UrQMD) model generated data to investigate the intermittent fluctuation of single-particle density distribution spectrum in the light of scaled factorial moments (SFM) for the minimum bias events of pp collisions at the LHC energies $\sqrt{s}$ = 13 TeV in the pseudo-rapidity (η), azimuthal angle (φ) and two dimensional (η − φ) phase space and to search for any possible existence of QGP phase transition. Study of fractal nature of particle production process in pp collisions in terms of UrQMD simulation will also be investigated. Before going into the details of the analysis it will be convenient for the readers to have brief introductions about the UrQMD model.

## UrQMD model

UrQMD model is a microscopic transport theory, based on the covariant propagation of all the hadrons on the classical trajectories in combination with stochastic binary scattering, colour string formation and resonance decay [22-24]. It represents a Monte Carlo solution of a large set of coupled partial integro-differential equations for the time evolution of various phase space densities. The basic input to this transport model is that, a hadron–hadron interaction would occur if the relative distance ($d_{trans}$) between the two particles in three-dimensional configuration space satisfies the relation $d_{trans} \leq d_0 = \sqrt{\frac{\sigma_{tot}}{\pi}}$ [26-28]. Total cross-section $\sigma_{tot}$ depends on the centre-of-mass energy, the species and the quantum number of the incoming particles. At the point of closest approach, this distance is purely transverse with regard to the relative velocity vector of the particles. In UrQMD model the Fermi gas model has been utilized to describe the projectile and the target nuclei, the initial momentum of each nucleon being distributed at random between zero and the local Thomas–Fermi momentum. Each nucleon is described by a Gaussian shaped density distribution, and the wave function for each nucleus is taken as a product of single nucleon Gaussian functions without taking into account the Slater determinant necessary for anti-symmetrization. In configuration space the centroids of the Gaussian functions are distributed at random within a sphere, and finite widths of these Gaussian functions result in a diffused surface region. At low and intermediate energies, typically $\sqrt{s}$ < 5 GeV, the phenomenology of hadronic physics has been described in terms of interactions between known hadrons and their resonances. At energies above 5 GeV the excitation of color strings and their subsequent fragmentation into hadrons dominate the particle production mechanism. The rescattering effects are also nicely implemented into the model.  The UrQMD collision term contains 55 different baryon species (including nucleon, delta  and hyperon resonances with masses up to 2.25 GeV/$c^2$) and 32 different meson species (including strange meson resonances), which are supplemented by their corresponding anti-particle and all isospin-projected states. The states can either be produced in string decays, s-channel collisions or resonance decays. This model can be used in the entire available range of energies from the Bevalac region to RHIC and LHC. For more details about this model, readers are requested to consult [22-24].

## Results and Discussions

In order to search for the possibility of existence of  quark-hadron phase transition during the charged particle production in  pp collisions at the LHC energies at $\sqrt{s}$ = 13 TeV by the method of scaled factorial moment we have generated a large sample of minimum-bias events  using the UrQMD model (UrQMD-3.3p1).  The intermittent pattern in the emission source of the particle in high-energy pp collisions may be different in different phase space, depending on the nature of the emission spectra of the charged particles. To find out nature of the intermittent pattern in pp collisions at the LHC energies, the analysis of intermittent type of fluctuations has been carried out both in one dimensional pseudo-rapidity (η) and azimuthal angle (ϕ) phase space.

Before going to the details of our analysis by the method of scaled factorial moment, we should remember that the particle distribution in the considered phase space is non–flat. This introduces some trivial fluctuations in the extracted dynamical fluctuations. According to the method enunciated by A. Bialas and M. Gazdzicki [25] we can get rid of this situation. Bialas and Gazdzicki [25] pointed out that one should replace the considered phase space variable with a cumulative variable $X_x$ defined as [25]
$$X_x = \frac{\int_{x_1}^{X} \rho(X) \partial X}{\int_{x_1}^{x_2} \rho(X) \partial X}$$
………………………………….(3)

Where $x_1$ and $x_2$ are the minimum and the maximum values of the concerned phase space variable. The variable $X_x$ varies between 0.0 to 1.0 so that the density distribution $\rho(X_x)$ remains almost constant [26]. We have transformed the pseudo-rapidity variable $\eta$ (-8 to+8) to the cumulative variable $X_\eta$ and the azimuthal angle ϕ (0-360 degree) to the cumulative variable $X_\phi$ using equation (3). The values of $X_\eta$ and $X_\phi$ lie between 0 to 1. Later in this paper whenever we mention the pseudo-rapidity space η or azimuthal angle space ϕ, it should be understood that we actually mean $X_\eta$ and $X_\phi$, the transformed variables having value lying between 0 and 1.

We have divided the transformed pseudo-rapidity and azimuthal angle space $X_\eta$ and $X_\phi$ respectively of the UrQMD simulated events of pp collisions at the LHC energies at $\sqrt{s}$ = 13 TeV in to M number of bins where M=2, 3, 4....20 so that $\delta\eta = \frac{X_\eta}{M}$ and also $\delta\phi = \frac{X_\phi}{M}$. We have calculated factorial moments $F_q$ for each bin according to the relation

$$F_q = M^{q-1} \sum_{m=1}^{M} \frac{\langle n_m(n_m-1)............(n_m-q+1)\rangle}{\langle n_m \rangle^q} \quad .............\ (4)$$ for both the phase spaces with q=2-5.

Here $n_m$ is the multiplicity of m th bin and q is the order of moment. $\langle ............ \rangle$ signifies the event average. The variation of ln<$F_q$> with lnM in case of UrQMD model simulated pp collisions at $\sqrt{s}$ = 13 TeV in pseudo-rapidity (η) and azimuthal angle (ϕ) space have been presented in figure 1(a) and in figure 1(b) respectively. Error bars drawn to each of the experimental points of the figures represent the statistical errors only. Statistical errors of the factorial moments have been evaluated from the standard deviations of different $F_q$ values of the event samples of the interaction. We have fitted the graphical points of all the plots with the best linear behavior and the slope of the best linear behavior give the intermittency exponent $\alpha_q$ of different order q for the said interaction. Calculated values of the intermittency exponents for both the phase spaces have been presented in the table 1. From the table it may be seen that the intermittency exponents increase with the order of moment q suggesting the presence of non-statistical fluctuations in simulated pp collisions at $\sqrt{s}$ = 13 TeV in both the phase spaces. Table 1 also reflect that the intermittent fluctuations in azimuthal angle space are quite weak in comparison to that of the pseudo-rapidity space.

We have also calculated the values of anomalous fractal dimension $d_q$ from the intermittency exponents using the relation $d_q = \frac{\alpha_q}{q-1}$. The anomalous fractal dimensions are found to increase with the increase of order q indicating the presence of multifractality in multiparticle production process for both the phase spaces which could reflect the presence of cascading mechanism of particles production in pp collisions. The variations of $d_q$ with order number q has been shown in figure 3 for UrQMD simulated data of pp collisions at $\sqrt{s}$ = 13 TeV. We have also calculated from the values of $\beta_q$ from the ratio of $\frac{d_q}{d_2}$ for the UrQMD simulated data in pseudo-rapidity (η) and azimuthal angle (ϕ) space. The calculated values of the $\beta_q$ have been presented in table 1 for UrQMD simulated data of pp collisions at $\sqrt{s}$ = 13 TeV.

In order to search for the Ginzburg-Landau phase transition by the method of factorial moment we have studied the variation of $\beta_q$ with (q-1) in figure 4 for UrQMD simulated pp collision data at $\sqrt{s}$ = 13 TeV in pseudo-rapidity (η) and azimuthal angle (ϕ) space. The variations of $\beta_q$ with (q-1) have been fitted with the function $\beta_q = (q-1)^\nu$ in order to extract the critical exponent υ. In table 1 we have shown the calculated values of the critical exponents for UrQMD simulated pp collision data at $\sqrt{s}$ = 13 TeV calculated from scaled factorial moment analysis in the pseudo-rapidity and azimuthal angle space. In pseudo-rapidity and azimuthal angle space the value of υ were found to be 1.472±.008 and 1.582±.011 respectively as evident from table 1. The critical exponents calculated by factorial moment methods are far away from the value 1.304 signifying the absence of quark-hadron phase transition in both the phase spaces.

## Two Dimensional Analysis

So far, we have searched for the existence of quark-hadron phase transition in UrQMD model simulated pp collisions at $\sqrt{s}$ = 13 TeV by the scaled factorial moment method in one-dimensional phase space. As the actual process takes place in three dimensions, the analysis should be carried out in higher dimensions instead of studying it in one dimension. Ochs [27] stressed the necessity of studying the fluctuation pattern in higher dimension. He pointed out that fluctuations in higher dimensions may not show up when projected on to one dimension. According to him, in lower dimension, the effects of fluctuation are reduced or they can even be completely washed out. A higher dimensional analysis will obviously provide more information on particle fluctuations. Therefore, a higher dimensional analysis will be highly appreciated. For a detail discussion readers may consult [4].

In two dimensions if we denote the two phase space variables as $x_1$ and $x_2$, the factorial moment of order $q$ can be defined as followed by Bialas and Peschanski [1]

$$F_q(\delta x_1, \delta x_2) = M^{q-1} \sum_{m=1}^{M} \frac{\langle n_m(n_m - 1)\ldots\ldots\ldots(n_m - q + 1)\rangle}{\langle n_m \rangle^q} \ldots\ldots\ldots\ldots(5)$$

Where $\delta x_1 \delta x_2$ is the size of a two dimensional cell. $n_m$ is the multiplicity of the m-th cell in this case.

Let us take a two dimensional region $\Delta x_1 \Delta x_2$ and divide it in two subshells of width $\delta x_1 = \frac{\Delta x_1}{M_1}$ and $\delta x_2 = \frac{\Delta x_2}{M_2}$. Where $M_1$ be the number of bins along $x_1$ direction and $M_2$ be the number of bins along $x_2$ direction. $M_1$ and, $M_2$ are the scale factors that satisfy the relation $M_1 = M_2^H$. $H$ is called the Hurst exponent [28] and it is given by $H = \frac{\ln M_1}{\ln M_2}$ with $M_1 \leq M_2$ and $0 \leq H \leq 1$. It is the parameter that characterizes the anisotropy of the system under study and fluctuations are self affine in nature if $H < 1$. For $H = 1$, $M_1 = M_2$, the system is isotropic in these two directions and self-similar fluctuations are on the cards.

The value of the Hurst exponent can be optimized by finding the minimum of $\chi^2$ per degrees of freedom ($\chi^2$/DOF) value of the linear fit to the plot of $\ln\langle F_q \rangle$ against $\ln M$. This procedure may not give the correct value of Hurst exponent always. Because in this case the corresponding errors should be estimated with utmost care keeping in mind that the data points are correlated since they stem from the same data sample which is quite difficult. Though different approaches for error calculation have been prescribed, none of them has been claimed to give a correct estimation. The estimation of errors is nontrivial and the accuracy of the method of determination of depends on how accurately the errors are determined. The robustness of this method of determination of H was questioned by some physicists. Another drawback of this method is that one cannot be sure about the correct partitioning condition. Let us explain it in details. The study of the multiplicity fluctuations in the self-affine space can be investigated by shrinking the phase space interval size in two anisotropic independent directions in two different ways. One can choose either $M_\eta$ or $M_\phi$ to be an integer. If one select $M_\phi$ to be an integer, $M_\eta$ will be related to $M_\phi$ as $M_\phi = M_\eta^H$. On the other hand if $M_\eta$ is chosen to be an integer, the relation between $M_\eta$ and $M_\phi$ will be $M_\eta = M_\phi^H$. The above discussion signifies that either $\eta$ space is divided in to finer intervals or the $\phi$ space is divided in to finer intervals. The optimization of Hurst exponent by calculating the minimum of ($\chi^2$/DOF) value of the linear fit to the plot of $\ln\langle F_q \rangle$ to $\ln M$ cannot select between $M_\eta = M_\phi^H$ and $M_\phi = M_\eta^H$. In order to overcome these difficulties and to be ascertained about the correct partitioning condition we have to look for an alternative method to find the Hurst exponent H.

According to Agababyan et al [29] that the Hurst exponent can be deduced from the data by fitting the one-dimensional second-order factorial moment saturation curves [29] against the number of bins (M) for both the phase spaces in the following manner $F_2^{(i)}(M_i) = A_i - BM_i^{-C_i}$ ........ (6)

The fitting parameter $C_i$ is related with the Hurst exponent $H_{ij}$ by the relation

$$H_{ij} = \frac{1+C_j}{1+C_i} \quad \ldots\ldots\ldots\ldots\ldots\ldots(7)$$

When $C_\eta > C_\phi$, $H = \frac{1+C_\phi}{1+C_\eta}$. $M_\eta$ will be related to $M_\phi$ by the relation $M_\eta = M_\phi^H$. On the other hand, if $C_\phi > C_\eta$ $H = \frac{1+C_\eta}{1+C_\phi}$ in that case the relation between $M_\eta$ and $M_\phi$ will be $M_\phi = M_\eta^H$. Thus, the value of the fitting parameter $C_i$ asserts the correct partitioning condition.

In this paper, we have repeated the two dimensional analysis of the scaled factorial moment method with $M$ varying from 2 to 20 for q=2-5. We have used equation (6) and (7) to find the values of the Hurst exponent for the said interaction. In order to extract the value of Hurst exponent, we have studied the variation of second order factorial moment $F_2$ against the number of bins M for UrQMD model simulated pp collisions at $\sqrt{s}$ = 13 TeV in both the phase spaces under the binning condition $2 \leq M \leq 20$. The plot of $F_2$ against M is shown in fig 2(a) and 2(b) for UrQMD model simulated pp collisions at $\sqrt{s}$ = 13 TeV in pseudo-rapidity and azimuthal angle phase spaces respectively. The values of A, B and C as obtained from the curve fitting analysis has been shown in table 2. From the table 2 we see that $C_\eta > C_\phi$ for UrQMD model simulated pp collisions at $\sqrt{s}$ = 13 TeV. Thus, the partitioning condition for UrQMD model simulated pp collisions at $\sqrt{s}$ = 13 TeV is $M_\eta = M_\phi^H$. The value of Hurst exponent comes out to be 0.955. Now we have calculated the values of normalized factorial moment for UrQMD model simulated pp collisions at $\sqrt{s}$ = 13 TeV under the partitioning condition $M_\eta = M_\phi^H$ with H=0.955 for q=2-5 in the two dimensional($\eta - \phi$) phase space. For this study the partition number along the $\phi$ direction stands as $M_\phi = 2,3\ldots\ldots 20$.

The values of the normalized factorial moments in the two-dimensional anisotropic ($\eta - \phi$) phase space under the correct partitioning condition have been calculated for UrQMD model simulated pp collisions at $\sqrt{s}$ = 13 TeV for q=2-5. The plot of ln<$F_q$> with lnM in case of the two-dimensional anisotropic ($\eta - \phi$) phase space has been shown in fig 1(c). Errors shown in figure 1(c) are nothing but statistical errors. As before we have fitted the experimental points of the plot ln<$F_q$> with lnM with a straight line and from the slope of the straight fit we have computed the strength of the intermittency exponent $\alpha_q$ and presented the values in table 1. From the table it is obvious to note that stronger fluctuations are present in the two dimensional anisotropic ($\eta - \phi$) phase space. We have also calculated the values of anomalous fractal dimension $d_q$ for different orders (q=2-5) and the values have been presented in table 1. Presence of multifractality is reflected from the order dependence of $d_q$ as evident from the table. The variation of $d_q$ on q for UrQMD model simulated pp collisions at $\sqrt{s}$ = 13 TeV in the two dimensional anisotropic ($\eta - \phi$) phase space has been depicted in figure 3. The presence of multifractality in two dimensional space also speaks in favour of the presence of cascading type of mechanism in the particle production process as in one dimensional space. We have also calculated $\beta_q$ from the ratio of $\frac{d_q}{d_2}$ for the UrQMD simulated data in the two-dimensional anisotropic ($\eta - \phi$) phase space using relation (1). The values of the $\beta_q$ calculated for the two dimensional space have also been presented in table 1 for UrQMD simulated data of pp collisions at $\sqrt{s}$ = 13 TeV. In order to search for the existence of GL phase transition in two-dimensional anisotropic ($\eta - \phi$) phase space for UrQMD simulated data of pp collisions at $\sqrt{s}$ = 13 TeV, variation of $\beta_q$ with q-1 has been shown in figure 4. From the best fitted curve the critical exponent $\upsilon$ has been calculated. In two-dimensional anisotropic (η − ϕ) phase space the value of υ has been found to be 1.743±.016 as presented in table 1 excluding any possibility of quark hadron phase transition.

## Conclusions

An event-by-event intermittency analysis is performed for the charged particle multiplicity distributions of the minimum bias events generated using UrQMD model in pp collisions at the LHC energies $\sqrt{s}$ = 13 TeV in the pseudo-rapidity (η), azimuthal angle (ϕ) and two dimensional anisotropic (η − ϕ) phase space. UrQMD simulated minimum bias events of pp collisions exhibit strong intermittency in the pseudo-rapidity phase space. However in comparison to that weak intermittency has been observed in case of azimuthal angle space. Stronger intermittent type of fluctuations has been observed in the two dimensional anisotropic (η − ϕ) phase space. Estimation of anomalous dimension $d_q$ and its variation with the order of the moment q suggests a multifractal nature of emission spectra of high-multiplicity pp events and is attributed to the particle production through cascading mechanism in the pseudo-rapidity (η), azimuthal angle (ϕ) and two dimensional anisotropic (η − ϕ) phase space. No evidence of quark-hadron phase transition is observed for the simulated pp collision data searched by factorial moment method in any phase space.

It should be mentioned here that so far bo such analysis have been performed with the experimental data of pp collisions at $\sqrt{s}$ = 13 TeV at the LHC. Such studies should be interesting to see whether we get similar observations from the experimental analysis..

**Table 1**

| Data | Phase space | q | $\alpha_q$ | $d_q$ | $\beta_q$ | $v$ |
|---|---|---|---|---|---|---|
| UrQMD simulated pp collisions at $\sqrt{S}$=13TeV | Pseudo-rapidity Space | 2 | 0.036±0.001 | 0.036±0.001 | 1.000±0.002 | 1.472±.008 |
| | | 3 | 0.096±0.004 | 0.048±0.002 | 2.670±0.004 | |
| | | 4 | 1.760±0.009 | 0.059±0.004 | 4.920±0.007 | |
| | | 5 | 2.820±0.016 | 0.070±0.010 | 7.780±0.009 | |
| UrQMD simulated pp collisions at $\sqrt{S}$=13TeV | Azimuthal angle Space | 2 | 0.008±0.001 | 0.008±0.001 | 1.000±0.002 | 1.582±.011 |
| | | 3 | 0.024±0.004 | 0.012±0.002 | 3.000±0.004 | |
| | | 4 | .047±0.006 | 0.0159±0.004 | 5.940±0.007 | |
| | | 5 | 0.071±0.008 | 0.0177±0.010 | 8.840±0.007 | |
| UrQMD simulated pp collisions at $\sqrt{S}$=13TeV | Two-dimensional phase space | 2 | .158±0.006 | 0.158±0.006 | 1.000±0.005 | 1.743±.016 |
| | | 3 | .647±0.022 | 0.012±0.002 | 4.080±0.004 | |
| | | 4 | 1.24±0.025 | 0.0159±0.004 | 7.830±0.009 | |
| | | 5 | 1.61±0.082 | 0.0177±0.010 | 10.600±0.011 | |

Table1 represents the values of $\alpha_q$, $d_q$, $\beta_q$ for different orders and the critical exponent $v$ calculated by normalized Factorial moment method in case of UrQMD simulated pp collision data at $\sqrt{s}$ = 13 TeV for pseudo-rapidity ,azimuthal angle and two dimensional anisotropic phase space .

**Table 2**

| Interactions | Phase space | A | B | C | H | Partitioning Condition |
|---|---|---|---|---|---|---|
| pp collisions at $\sqrt{s}$=13 TeV | Pseudo-rapidity | 1.72±.11 | .341±.010 | 0.272±.011 | 0.955 | $M_\eta = M_\phi^H$ |
| | Azimuthal angle | 1.47±.13 | .077±.11 | 0.216±.012 | | |

Table 2 represents the values of the fitting parameters A, B, C of the equation $F_2^{(i)}(M_i) = A_i - BM_i^{-C_i}$, the Hurst exponent H and the partitioning condition in case of UrQMD simulated pp collision data at $\sqrt{s}$ = 13 TeV obtained from the curve fitting analysis of $F_2$ against M in both pseudo-rapidity and azimuthal angle phase space .

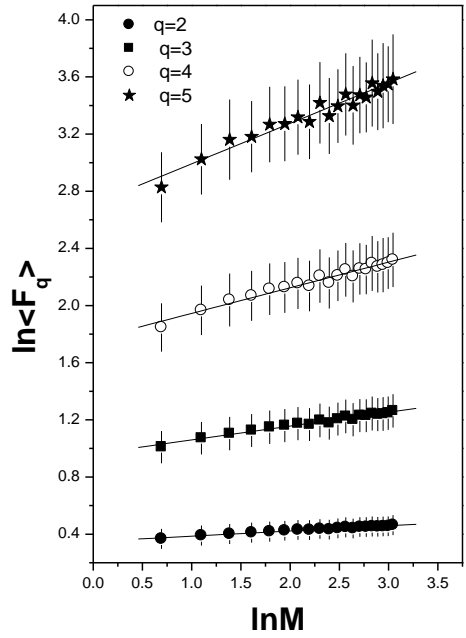

Figure 1(a)  The variation of ln<$F_q$> with lnM in case of  UrQMD model simulated pp collisions at $\sqrt{s}$ = 13 TeV in pseudo-rapidity (η) space.

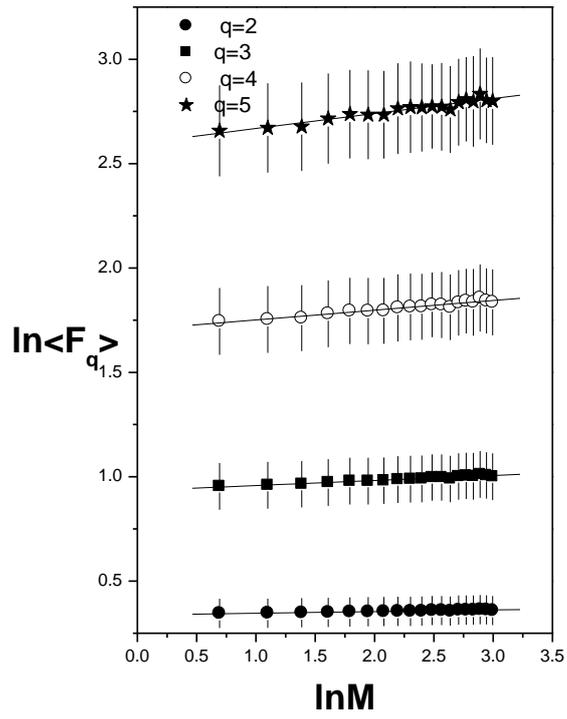

Figure 1(b)   The variation of ln<$F_q$> with lnM in case of UrQMD model simulated pp collisions at $\sqrt{s}$ = 13 TeV in azimuthal angle space.

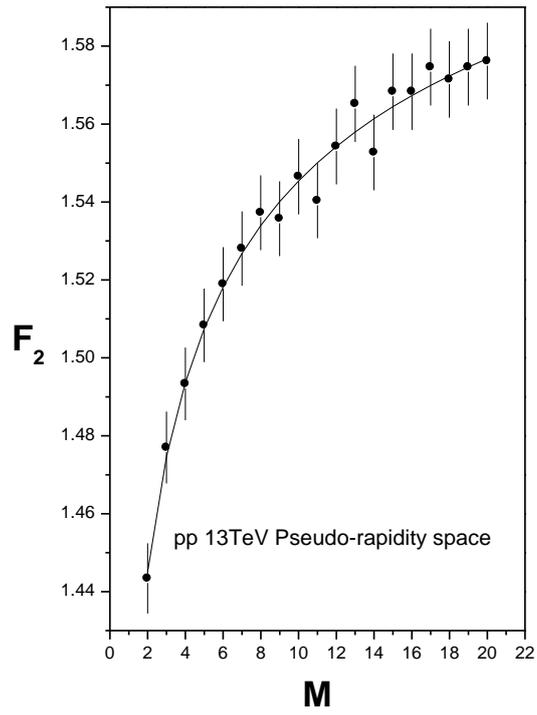

Fig 2(a) represents the plot of $F_2$ against $M$ for pseudo-rapidity space in simulated pp collisions at $\sqrt{s}$=13 TeV

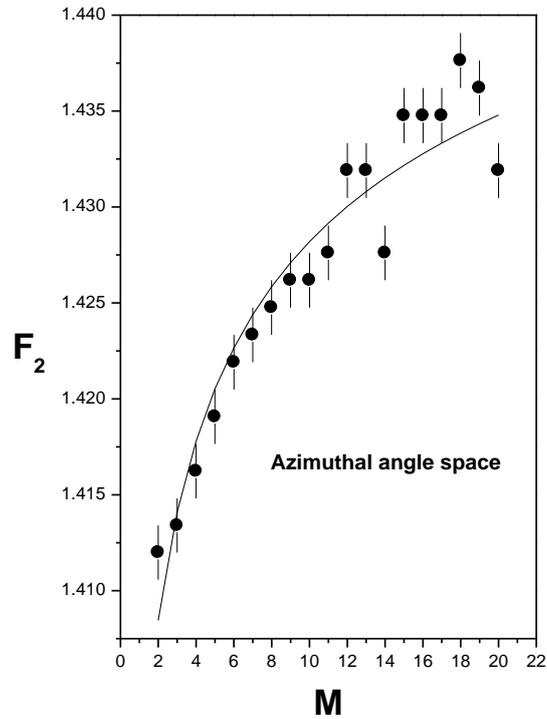

Fig 2(b) represents the plot of $F_2$ against $M$ for azimuthal angle space simulated pp collisions at $\sqrt{s}$=13 TeV

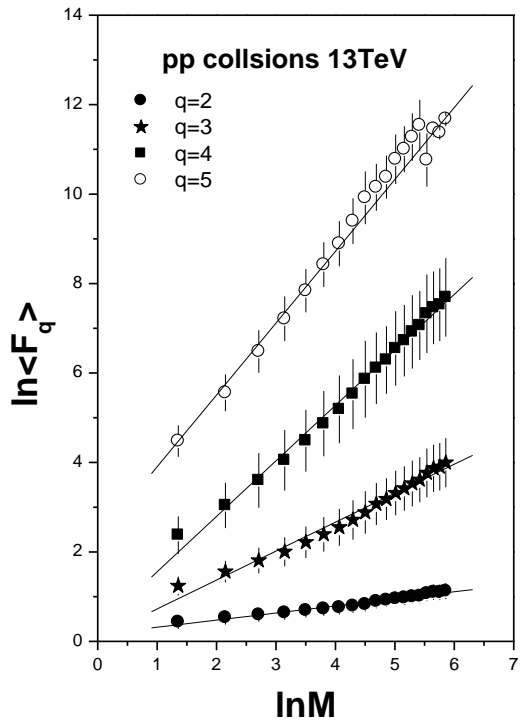

Figure 1 (c)  The variation of ln<$F_q$> with lnM in case of  UrQMD model simulated pp collisions at $\sqrt{s}$ = 13 TeV in  two dimensional phase space.

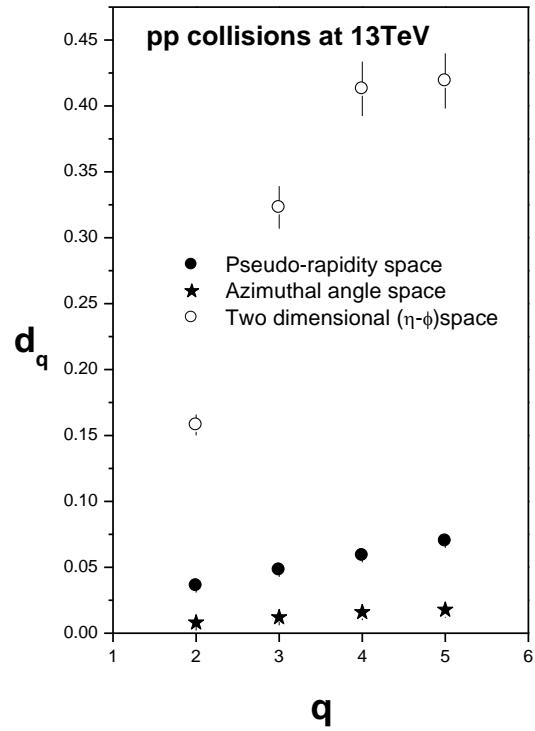

Figure 3 Variation of $d_q$ with order q in case of UrQMD simulated pp collisions at $\sqrt{s}$=13 TeV in pseudo-rapidity, azimuthal angle and two dimensional anisotropic phase space.

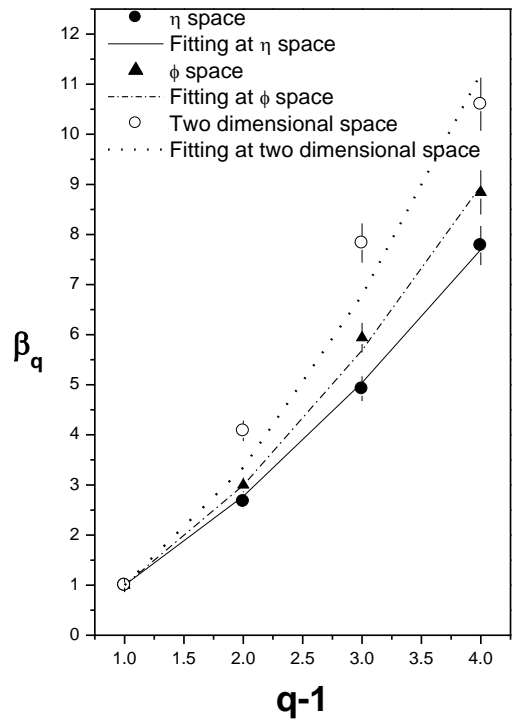

Figure 4 represents the fitting of $\beta_q = (q-1)^\nu$ in case of UrQMD simulated pp collisions at $\sqrt{s}$=13 TeV in pseudo-rapidity ,azimuthal angle and two dimensional anisotropic phase space.